\documentclass[aps, showpacs, pra,twocolumn, a4paper]{revtex4-1}
\usepackage{latexsym}
\usepackage{graphics}
\usepackage{amsmath,amssymb}
\usepackage{mathcomp}
\usepackage{array}
\usepackage{verbatim}
\usepackage{float}

\usepackage[normalem]{ulem} 

\newcommand{\pAHz}{\ensuremath{\mathrm{pA}/\sqrt{\Hz}}}

\newcommand{\Tesla}{\ensuremath{\mathrm{T}}}

\newcommand{\fTHz}{\ensuremath{\fT/\sqrt{\Hz}}}

\newcommand{\fT}{\ensuremath{\mathrm{f}\Tesla}}

\newcommand{\Hz}{\ensuremath{\mathrm{Hz}}}

\newcommand{\polplus}{\ensuremath{\sigma_{+}}}
\newcommand{\polminus}{\ensuremath{\sigma_{-}}}

\begin{document}
\title{A high-sensitivity push-pull magnetometer }

\author{E. Breschi,
Z. D. Gruji\'c, P. Knowles, and A. Weis}
%
%
\affiliation{Department of Physics, University of Fribourg, Fribourg 1700, Switzerland}
%

\begin{abstract}
We describe our approach to atomic magnetometry based on the push-pull optical pumping technique.
Cesium vapor is pumped and probed by a resonant laser beam whose circular polarization is modulated synchronously with the spin evolution dynamics induced by a static magnetic field.
The magnetometer is operated in a phase-locked loop, and it has an intrinsic sensitivity below 20\fTHz{} using a room temperature paraffin-coated cell.
We use the magnetometer to monitor magnetic field fluctuations with a sensitivity of 300\fTHz{}.
\end{abstract}

\maketitle

A scalar atomic magnetometer measures the modulus of a static magnetic field via the Larmor frequency at which the atomic magnetic moments precess coherently.
Resonant laser light is used both to create a macroscopic magnetization by orienting the atomic spins,
 and to detect the effect of the precessing magnetization on the medium's optical absorption coefficient.

Atomic magnetometry dates back to the 1960s
\cite{Bell:1961:ODS} and in the 1990s, interest in the
topic resurged due to the development of compact diode lasers
and microfabrication technologies.
Several recent review articles have been devoted to atomic
magnetometry
\cite{Budker:2002:RNM,Alexandrov:20xx:RAA,Budker:2007:OM}.

In a traditional atomic magnetometer, polarized light resonant with an atomic absorption line
produces an imbalance of magnetic sublevel populations by optical
pumping, thus creating spin polarization, and an associated macroscopic magnetization.
In the so-called double resonance magnetometer (which may be realized in $M_x$- or $M_z$- configuration, see \cite{Alexandrov:20xx:RAA}), a weak  magnetic field, referred to as radio-frequency or `rf' field,
oscillating at frequency $\nu_{rf}$, drives transitions between neighboring Zeeman-split sublevels, thereby
destroying the polarization, an effect that is resonantly enhanced when $\nu_{rf}$ matches the Larmor frequency, $\nu_L$.
This principle finds a widespread use in commercial magnetometers.

One may view the rf field in the scheme outlined above as a mechanism that synchronizes the spin precession of the polarized atoms.
In an alternative approach, spin synchronization is achieved by a suitable modulation
of the pumping light \cite{Grujic:2013:AMR}
using amplitude~\cite{Schultze:2012:CPI}, frequency~\cite{Budker:2002:NMO,Belfi:2007:AOS,Pollinger:2012:EIO}, or
polarization~\cite{Ben-Kish:2010:DZF,Fescenko:2013:BBE,Breschi:2013:MOS} modulation.
The latter approaches to magnetometry yield \emph{magnetically-silent}
magnetometers, in which no oscillating magnetic field is applied to the sensor proper.
The sensor thus does not produce any field other than the excessively weak field of the polarized atoms themselves.
This is an important aspect for avoiding sensor crosstalk in multi-sensor applications.

Here we present a so-called push-pull magnetometer that is based on the
modulation - at the Larmor frequency - of the light beam's polarization between left- and right-circular.
The original proposal of the push-pull optical pumping technique \cite{Jau:2004:PPO} aimed at increasing the contrast of the magnetically insensitive transitions in atomic clocks
by polarization modulation at the clock (i.e., hyperfine transition) frequency.
The method has been demonstrated for the clock
transition in rubidium \cite{Jau:2005:SSF}, potassium
\cite{Jau:2007:PPL}, and cesium \cite{Liu:2013:CPT}.
So far, it has not been explored in the case of magnetically
sensitive resonances.
The concept of push-pull (`pp') refers to the populations of atomic
sublevels being pushed and pulled between specific magnetic sublevels by the
interaction with laser radiation whose polarization is
modulated at the frequency of the coherent evolution of the quantum superposition of those states.
In microwave-pp the sublevel dynamics is driven by the hyperfine interaction, while in Zeeman-pp (relevant here) it is driven by the static magnetic field.

In our experiments, the magnetization of a spin-oriented medium prepared by optical pumping with circularly polarized laser light, precesses around a static magnetic field that is perpendicular to the
light propagation direction $\vec{k}$.
After half a Larmor period, an initial spin polarization prepared, say  by
$\polplus$-pumping reverses its sign.
At the time when the spin is fully reversed, the light polarization will be $\polminus$ and will thus further increase the spin polarization.
This  process repeats periodically, until a steady-state precessing polarization of constant amplitude is reached.
The light polarization switching at the Larmor frequency will thus efficiently preserve the overall (precessing) atomic
spin polarization yielding high contrast and narrow resonance signals.


\begin{figure}
\centerline{\resizebox{1\columnwidth}{!}{\includegraphics{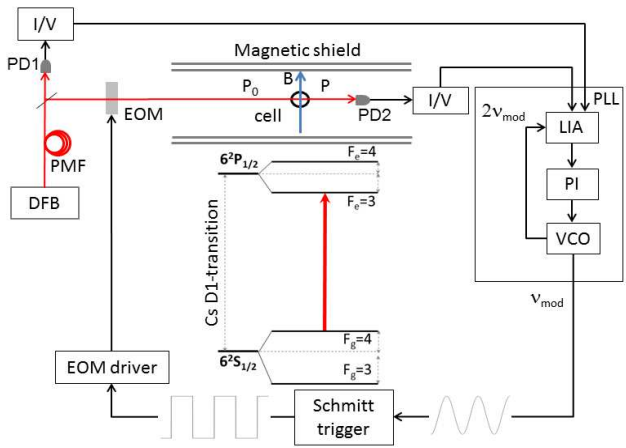}}}
  \caption{Experimental apparatus. $P_0$ is the incident laser power.
    DFB: distributed feedback diode laser, PMF: polarization maintaining optical fiber, EOM:   electro-optic modulator,  PD1,PD2: non-magnetic photodiodes, I/V:
    current-voltage converter,  PLL: phase-locked-loop, LIA: lock-in amplifier, and VCO: voltage-controlled oscillator. }
  \label{Fig:scheme}
\end{figure}

Figure~\ref{Fig:scheme} shows a block-diagram of the experimental
apparatus.
We use a DFB laser emitting 894~nm radiation near the cesium $6^{2}S_{1/2}\rightarrow 6^2P_{1/2}$ transition (energy structure shown as inset).
The laser frequency is actively stabilized to the $F_g{=}4\rightarrow F_e{=}3$ transition that is known to yield the highest contrast in
 magneto-optical spectroscopy.
The optical part of the experimental apparatus has been described in
detail in Ref.~\cite{Fescenko:2013:BBE}, while the magnetic field
generation and control are addressed in Ref.~\cite{Breschi:2013:SCM}.

The set-up is mounted inside of two $\approx$~1200~mm long mu-metal cylinders (diameters of 320 and ~290~mm, respectively) without endcaps, whose axes are orthogonal to the local laboratory field.
We have shown that the residual magnetic field in our shield
can be compensated at the nT level~\cite{Breschi:2013:SCM}.

The laser beam's polarization is modulated by a commercial electro-optical
modulator (EOM) with a square-wave of $50\%$ duty cycle between left($\polplus$)- and right($\polminus$)-circular polarization states.
The polarization modulation frequency, $\nu_{mod}$, is swept around the resonant
value $\nu_{L}$.

The spectroscopy cell is a $30$~mm diameter evacuated paraffin-coated
glass bulb~\cite{Castagna:2011:MLT} containing cesium vapor at $\sim$20$^\circ$~C.
The light transmitted through the cell is detected by a photodiode (PD2), whose photocurrent is amplified by a transimpedance amplifier and demodulated by a commercial
computer-controlled digital lock-in amplifier (Zurich Instruments, model HF2LI) referenced to the 2nd harmonic of $\nu_{mod}$.
The HF2LI allows for a differential input.
We make use of this by subtracting the properly amplified reference signal of the photodiode PD1 (Fig.~\ref{Fig:scheme}) from the PD2 signal, thereby suppressing common mode power fluctuations.

The lock-in's in-phase, quadrature or phase outputs can be used either for spectroscopic studies,
for signal characterization, or to operate the apparatus as a magnetometer using a phase-locked (PLL) loop.
In the latter case the lock-in serves as a phase detector, whose phase output, after proportional-integral amplification, drives a voltage-controlled oscillator (VCO).
A Schmitt trigger transforms the VCO-generated sine-wave into a square-wave that drives the EOM\@.
We note that the HF2LI instrument contains all necessary software components for realizing such a PLL.

The degree of polarization of the $\sigma_+/\sigma_-$ states was determined to deviate by less than 0.5~\% from perfect circular polarization.
A constant magnetic field is applied orthogonally to the laser
propagation vector, and $\nu_{mod}$   is scanned around the strongest magneto-optical resonance (occurring at occurring at $\nu_{mod}=\nu_L$~\cite{Breschi:2013:SCM}) in order to adjust the PPL parameters.
We note that this type of magnetometer has a dead zone when the magnetic field is aligned along the laser beam.

\begin{figure}
\centerline{\resizebox{1\columnwidth}{!}{
  \includegraphics{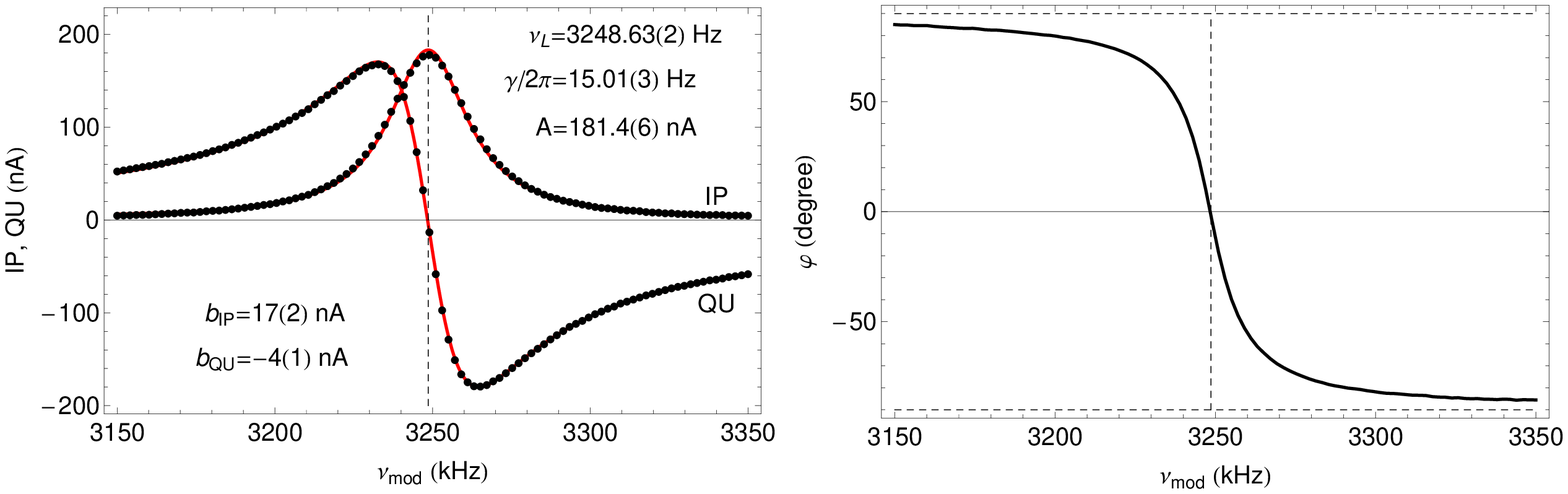}}}
  \caption{In-phase and quadrature
  signals (left), and the phase (right) of the differential photodetector signal (PD2-PD1), demodulated at $2\,\nu_{mod}$.
  The resonance occurs when $\nu_{mod}=\nu_{L}$.}
  \label{Fig:Sgn}
\end{figure}

The left graph of Fig.~\ref{Fig:Sgn} shows the in-phase and
quadrature signals for an
incident laser power of 10~$\mu$W and a beam diameter of
4~mm.
The points represent experimental data while the solid lines are the
result of fits to the absorptive and dispersive Lorentzians predicted by the model \cite{Grujic:2013:AMR}
\begin{subequations}
\begin{align}
IP  &= b_{IP} + A\,\frac{\gamma^2}{(\omega_{mod}-\omega_L)^2+\gamma^2}\equiv  b_{IP} + A\,\frac{1}{1+x^2} \label{eqn:LorA}\\
QU  &= b_{QU} + 2\,A\,\frac{\gamma (\omega_{L}-\omega_{mod})}{(\omega_{mod}-\omega_L)^2+\gamma^2}\equiv b_{QU} + 2\,A\,\frac{x}{1+x^2} \,,
\label{eqn:LorD}
\end{align}
\end{subequations}
where $x=(\omega_{L}-\omega_{mod})/\gamma$, $A$ and $\gamma$ are the resonance amplitude and half-width.
The best-fit parameters are included in the graphs.
In Eqs.~(\ref{eqn:LorA}) and (\ref{eqn:LorD}), we have included
small possible background $b_{IP}$ and $b_{QU}$,and we will discuss their influence on the magnetometer performance below.

The right graph of Fig.~\ref{Fig:Sgn} shows the phase of the photocurrent's modulation with respect to the EOM-drive, calculated from $\varphi=\arctan (QU/IP)$, together with a fit.
The linear slope of the $\varphi(\nu_{mod})$ dependence near resonance serves as the discriminator in the PLL loop.
Note that neither the in-phase nor the quadrature signals in an ideal push-pull magnetometer have intrinsic backgrounds ($b_{IP}=b_{QU}=0$),
 a distinct advantage compared to amplitude (AM) or frequency (FM) modulation magnetometry methods.

A Taylor expansion shows that the phase signal near resonance is given by
\begin{equation}
\varphi(x)\approx\varepsilon_{QU}+2\,(1-\varepsilon_{IP})\,x \,,
\label{eqn:FiNearRes}
\end{equation}
to first order in the parameters $\varepsilon_{IP,QU}=b_{IP,QU}/A$ that describe the relative background contamination of the signals.
The last expression shows that a PLL locking the magnetometer to $\varphi=0$ will generate a systematic frequency error $\Delta\omega_{rf}\approx \gamma\,\varepsilon_{QU}/2$.
It also shows that the discriminator slope of the pp-magnetometer is $d\varphi/dx\approx 2$ when $\varepsilon_{IP}=0$, while a standard double resonance magnetometer, such as the $M_x$ magnetometer has a slope $d\varphi/dx=1$
\cite{Groeger:2006:HSL}.
The fit to the resonances of Fig.~\ref{Fig:Sgn} has revealed in fact
background levels of $\varepsilon_{IP}$=9\% and $\varepsilon_{QU}$=1\%
due to residual amplitude modulation.
The effective phase discriminator slope is then reduced to $d\varphi/dx\approx 1.8$.

Let us now estimate the sensitivity of the magnetometer.
For practical purposes we express all signals in terms of the
photocurrent that they generate.
In the expressions below, $P$ will therefore refer to laser power,
but will be expressed in current units.
Under ideal conditions the magnetometer performance will be limited
by the \textit{rms} shot noise $\sqrt{2\,e\,P_{DC}}$ of the average (DC) power
$P_{DC}$ detected by the photodiode in a 1 Hz bandwidth around the
modulation frequency $\nu_{mod}$, and $e$ is the electron charge.
We characterize the magnetometric sensitivity in terms of the
noise-equivalent magnetic field, $NEM$, i.e., the magnetic field
fluctuation $\delta B_{NEM}$ that yields phase noise equal
to the noise, $\delta P$, produced by power fluctuations.
A straightforward calculation shows that
\begin{subequations}
\begin{align}
\delta
B_{NEM}&=\frac{\gamma}{2\,\gamma_F}\,\frac{1}{1-\varepsilon_{IP}}\,\frac{\delta P}{A}\label{eq:NEM} \\
&\approx \frac{\gamma(P)}{2\,\gamma_F}\,\frac{\sqrt{2\,e\,P}}
                       {A(P)}                       \,, \label{eq:sNEM}
\end{align}
\end{subequations}
where the amplitude $A$ of the in-phase signal is expressed in current units, and where $\gamma_F\approx$ 2$\pi \times $3.5~Hz/nT is the gyromagnetic ratio of the $F=4$ ground state.
In equation (\ref{eq:sNEM}) we explicitly indicate that both the
resonance amplitude $A$ and width $\gamma$ depend on the laser power
$P$, and that the DC photocurrent $I_{DC}$ is equivalent to $P$.


\begin{figure}
  \centerline{\resizebox{1\columnwidth}{!}{
      \includegraphics{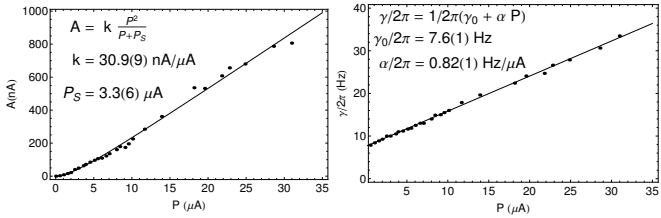}}}
  \caption{Laser power dependence of the resonance amplitude, $A$, and linewidth, $\gamma$.
 The laser power is expressed in terms of the DC photocurrent detected after the cell.}
  \label{Fig:VsP}
\end{figure}

In order to get a realistic estimate of the NEM that can be expected
in the shot noise limit we have measured the dependencies $A(P)$ and
$\gamma(P)$.
The results are shown in Fig.~\ref{Fig:VsP}, where the fit functions
and fitted parameters are included in the graphs.
One sees that at low power the signal amplitude grows quadratically with power, followed by a
linear dependence when saturation of the spin polarization sets in, and that the zero-power linewidth of $\gamma_0=2\pi \times $ 7.6(1)~Hz grows
linearly with power over the whole range of powers investigated.

\begin{figure}
\centerline{\resizebox{0.6\columnwidth}{!}{
  \includegraphics{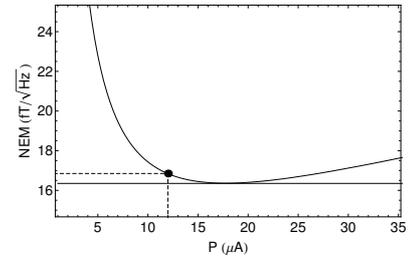}}}
  \caption{Shot noise-equivalent magnetic field (NEM) as a function of laser power, assuming no backgrounds, $\varepsilon_{IP}=\varepsilon_{QU}=0$.
The black dot represents the working point for the magnetometry measurements.
  }
  \label{Fig:NEM}
\end{figure}
Inserting the fitted functions of the experimental $A(P)$ and
$\gamma(P)$ dependencies into Eq.~\eqref{eq:NEM}, we obtain the
$\delta B_{NEM}(P)$  dependence shown in Fig.~\ref{Fig:NEM}.
We thus expect an optimal sensitivity below $\sim$17~\fTHz{} for DC
photocurrents in the range of $\sim$10--25~$\mu$A, corresponding to
$\sim$20--50~$\mu$W.
This sensitivity is comparable to the sensitivities that we have
observed in our lab~\cite{Castagna:2011:MLT} with paraffin-coated
cells of the same diameter using the $M_x$ magnetometer
technique.

\begin{figure}[ht]
\centerline{\resizebox{1\columnwidth}{!}{\includegraphics{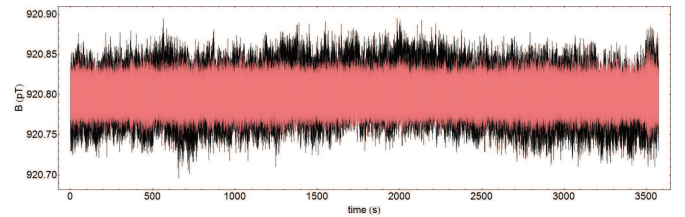}}}
\caption{One hour long recording of a nominally
constant magnetic field with the push-pull magnetometer operated
with PLL feedback.
Black: Raw PLL frequencies converted to field units; Pink: Same data after notch-filtering at 0, 50, and 150~Hz. }
\label{Fig:cMF}
\end{figure}

We have used the pp-magnetometer to monitor the magnetic field
variations inside of the two-layer shield.
For these measurements we operated the magnetometer in the PLL mode
shown in Fig.~\ref{Fig:scheme} with a power of $\sim$24~$\mu$W
($\sim$~12~$\mu$A), for which one expects a NEM of $\sim$17~\fTHz{} as
indicated by the black dot in Fig.~\ref{Fig:NEM}.
Figure~\ref{Fig:cMF} shows a time series of field values inferred
from the PLL frequency recorded at a rate of 450~samples/second with
a PLL-frequency resolution of 0.7$\mu$Hz, corresponding to 0.2~fT.

A Fourier analysis of these data reveals that they show, besides the
(dominant) slow drift visible in Fig.~\ref{Fig:cMF}, strong
oscillations at 50~Hz and 150~Hz.
The superimposed low noise trace in Fig.~\ref{Fig:cMF} shows the same data after digital
filtering with a series of notch filters (with a -3dB-bandwidth of
12.5~Hz) centered at 0, 50, and 150~Hz.

\begin{figure}
\centerline{\resizebox{.9\columnwidth}{!}{
  \includegraphics{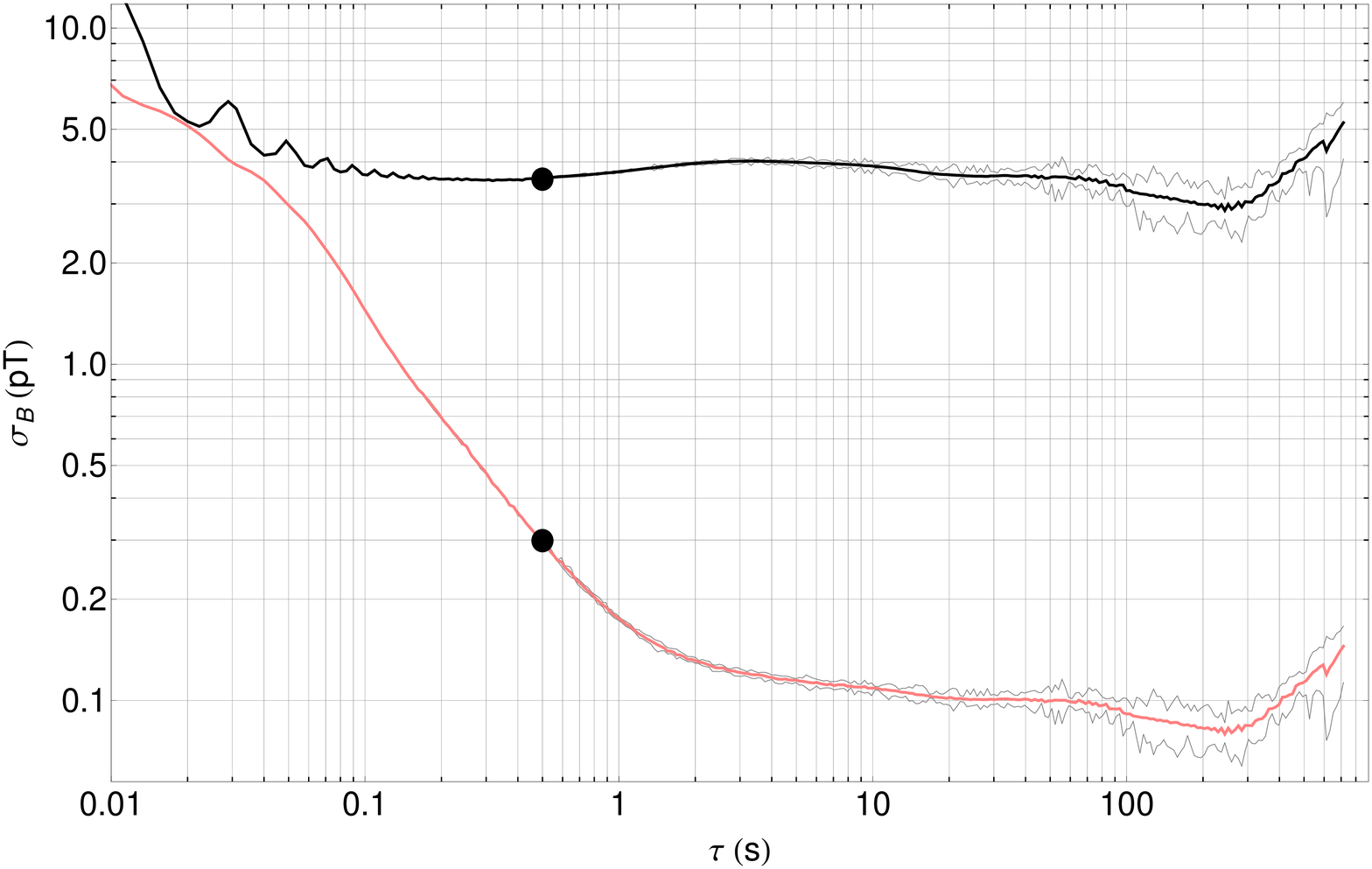}}}
  \caption{Allan standard deviation of the measured field as a
    function of the integration time, $\tau$ for the raw (black, top) and filtered (pink, bottom) data of Fig.~\ref{Fig:cMF}.}
  \label{Fig:Allan}
\end{figure}
Figure \ref{Fig:Allan} shows
the Allan standard deviation $\sigma_B(\tau)$ \cite{Allan66} of the raw (black, top) and filtered (pink, bottom) data of Fig.~\ref{Fig:cMF}.
The black dots at 3.6~pT and 300~fT, respectively, mark the $\sigma_B$-values for an integration time $\tau$ of 0.5~s corresponding to a 1~Hz bandwidth.

\begin{figure}
\centerline{\resizebox{1.\columnwidth}{!}{
  \includegraphics{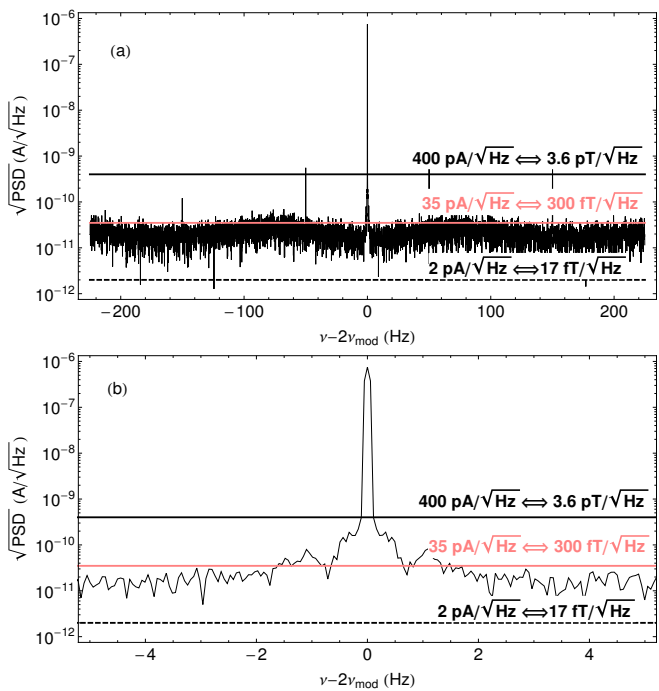}}}
  \caption{Square root of the photocurrent's power spectral density near the demodulation frequency of 2$\nu_{mod}$.
  The lower dashed line represent the shot noise of the DC photocurrent of 12$\mu$A.
  The upper two solid lines represent the current fluctuations inferred from the measured field fluctuations.}
  \label{Fig:ZoomFFT}
\end{figure}
 In Figs.~\ref{Fig:ZoomFFT}  we show zooms into the Fourier spectrum of the PLL input signal (Fig.~\ref{Fig:scheme})
in the ranges of $ \pm$225~Hz (Figs.~\ref{Fig:ZoomFFT}.a) and $\pm$5~Hz
(Figs.~\ref{Fig:ZoomFFT}.b), respectively, around the demodulation frequency of 2$\nu_{mod}$.
The peak at $\nu=2\nu_{mod}$ represents the PLL oscillation at the double Larmor
frequency.
It lies $\approx$20 times above the theoretical shot noise level of $\sim$2~\pAHz{} (equivalent to $\sim$17~\fTHz{}) of the $\sim$12$\mu$A photocurrent (lower dashed line).
The peak is superposed on a pedestal (well seen in Fig.~\ref{Fig:ZoomFFT}.b) that we
assign to slow magnetic field (and hence Larmor frequency)
fluctuations.
The pedestal stabilizes to a rather white noise floor of a few 10~pA/$\sqrt{Hz}$.

In order to get a better interpretation of these results we
calculate the  power noise levels  $\delta P$ (in a 1~Hz bandwidth)
that would yield the experimentally determined magnetic fluctuations
$\delta B=\sigma_B$(0.5~s) shown as black dots in Fig.~\ref{Fig:Allan}.
For this we solve Eq.~\eqref{eq:NEM} for $\delta P$ yielding
\begin{equation}\label{eq:deltaP}
    \delta P=\frac{2\,\gamma_F\,A(P)}{\gamma(P)}\,\delta B\,.
\end{equation}
Inserting $\delta B=$300~fT and 3.6~pT yields, for $P=12~\mu$A,
$\delta P\approx$35~ pA/$\sqrt{Hz}$ and $\delta
P\approx$400~pA/$\sqrt{Hz}$, respectively.
We show these values as horizontal lines in Figs.~\ref{Fig:ZoomFFT}.
The value $\delta P\approx$35~ pA/$\sqrt{Hz}$ inferred from the
Allan plot of the unfiltered data coincides well with amplitude of
the pedestal underlying the Larmor peak in the Fourier spectrum.
The value $\delta P\approx$400~ pA/$\sqrt{Hz}$, inferred from the
Allan plot of the notch-filtered time series (in which low frequency
drift and oscillations at the line frequency harmonics were removed)
coincides well with the quasi-white noise outside of the pedestal.
This comparison illustrates the internal consistency of our analysis, and in particular the validity of
Eq.\eqref{eq:NEM}.

The low frequency pedestal is most likely due to slow field drifts, be they from an instability of the current source or the slowly varying laboratory field that penetrates our open cylindrical shield.
Note that the  3.6~pT value at the pedestal's peak corresponds to a $\delta B/B$-variation of the 1~$\mu$T field of 3.6$\times$10$^{-6}$, demonstrating the high stability of our current source.
We have verified that the peaks at the line frequency and harmonics thereof, as well as the white noise floor of
$\approx~35~\pAHz$, originate from the variations of the DFB laser's power, frequency and phase.

The shot noise limited sensitivity of 17~\fTHz{} can thus only be
demonstrated experimentally with a more stable laser and current source or in a
gradiometer arrangement of magnetometers.



In conclusion, we have described an atomic magnetometer based on a
push-pull technique with polarization-modulated laser
light in a room temperature paraffin-coated Cs vapor cell.
The device, operated as a PLL, has an ultimate shot noise limited
sensitivity below 20\fTHz{} with a laser power around
10$\mu$W that is comparable to the performance of $M_x$
magnetometers operated with similar vapor cells.
The magnetometry method demonstrated here has
advantages compared to related magnetically-silent methods, such as
magnetometers based on FM- or AM-modulation.
Its background-free in-phase and quadrature signals make the pp-magnetometer performance less sensitive to not optimal lock-in phase settings.
Moreover, detection at the second harmonic, $2\,\omega_{mod}$,
 reduces contributions from spurious signals at the modulation frequency and suppresses noise contributions from $1/f$-noise.
Further investigations, such as a direct comparison of different
magnetometer methods deployed with the same cell in the same
experimental set-up under identical conditions, or the operation of
the pp-magnetometer in gradiometer mode, are foreseen.

\paragraph{acknowledgement}
This work is supported by the Ambizione grant $PZ00P2\_131926$ of the
Swiss National Science Foundation.
%
%
%

%
\end{document}